\documentstyle[sprocl,epsf]{article}

\def\b{\begin{eqnarray*}}
\def\e{\end{eqnarray*}}
\def\lsim {~^{<~}_{\sim~}}
\def\gsim {~^{>~}_{\sim~}}
\begin{document}
\title{CONFINEMENT PHYSICS AND TOPOLOGY IN QCD}
\author{H.~SUGANUMA, K.~AMEMIYA, A.~TANAKA and H.~ICHIE}
\address{Research Center for Nuclear Physics (RCNP), Osaka University\\
Mihogaoka 10-1, Ibaraki, Osaka 567-0047, Japan \\
E-mail: suganuma@rcnp.osaka-u.ac.jp} 

\maketitle\abstracts{
Based on the dual-superconductor picture, we study the confinement physics 
in QCD in terms of the monopole in the maximally abelian (MA) gauge using 
the SU(2) lattice QCD. In the MA gauge, the off-diagonal gluon component is 
forced to be small, and hence microscopic abelian dominance on the link 
variable is observed in the lattice QCD for the whole region of $\beta $. 
>From the gluon-propagator analysis in the lattice QCD, the origin of abelian 
dominance for the long-range physics is interpreted as the effective mass 
$m_{ch} \simeq 0.9 {\rm GeV}$ of the charged gluon induced by the MA gauge 
fixing. In the MA gauge, there appears the macroscopic network of the 
monopole world-line covering the whole system, which would be identified as 
monopole condensation at a large scale. Using the dual Wilson loop in the MA 
gauge, we find the effective mass of the dual gluon field, 
$m_B \simeq $0.5GeV, which is the evidence of the dual Higgs mechanism by 
monopole condensation. The large fluctuation of off-diagonal gluons remains 
around the monopole in the MA gauge, and this charged-gluon rich region 
would provide the effective monopole size as the critical scale of the 
abelian projected QCD. Instantons are expected to appear in the 
charged-gluon rich region around the monopole world-line in the MA gauge, 
which leads to the local correlation between monopoles and instantons.
}

\section{Dual Superconductor Picture for Confinement in QCD}
Quantum Chromodynamics (QCD) shows the asymptotic freedom 
due to its nonabelian nature, and the perturbative calculation 
is workable in the ultraviolet region as $\mu \gg$ 1GeV. 
On the other hand, in the infrared region as $\mu \lsim$ 1GeV, 
QCD exhibits nonperturbative phenomena 
like color confinement and dynamical chiral-symmetry breaking 
(D$\chi $SB) due to the strong-coupling nature.
Up to now, there is no systematic promising method 
for the study of the nonperturbative QCD (NP-QCD) except for the 
lattice QCD calculation with the Monte Carlo method.

The lattice QCD simulation can be regarded as a numerical experiment 
based on the rigid field-theoretical framework, 
and provides a powerful technique for the analyses of 
the NP-QCD vacuum and hadrons. 
Owing to the great progress of the computer power in these days, 
the lattice QCD becomes useful not only for ``reproductions of 
hadron properties'' but also for ``new predictions'' like 
glueball properties and the QCD phase transition.
The ``physical understanding'' for NP-QCD is also 
one of the most important subjects in the lattice QCD, 
because the aim of theoretical physics is never to reproduce numbers 
but to find the ``physical mechanism'' hidden in phenomena ! 
In this paper, we study the confinement physics in QCD 
using both the lattice QCD and the theoretical formalism.

In 1974, Nambu proposed 
an interesting idea that quark confinement would be 
physically interpreted using 
the {\it dual version of the superconductivity.}$^1$ 
In the ordinary superconductor, 
Cooper-pair condensation leads to the Meissner effect, 
and the magnetic flux is excluded or squeezed like a 
quasi-one-dimensional tube as the Abrikosov vortex, 
where the magnetic flux is quantized topologically. 
On the other hand, 
>from the Regge trajectory of hadrons and the lattice QCD results, 
the confinement force between the color-electric charge is 
characterized by the universal physical quantity of the string tension, 
and is brought by {\it one-dimensional squeezing} of the 
color-electric flux in the QCD vacuum. 
Hence, the QCD vacuum can be regarded as the dual version 
of the superconductor based on above similarities 
on the low-dimensionalization of the quantized flux between charges. 

In the dual-superconductor picture for the QCD vacuum, 
the squeezing of the color-electric flux between quarks 
is realized by the {\it dual Meissner effect,} 
as the result of {\it condensation of color-magnetic monopoles,} 
which is the dual version of the electric charge as the Cooper pair.
However, there are {\it two large gaps} between QCD and the 
dual-superconductor picture.

(1) 
This picture is based on the {\it abelian gauge theory} subject to the 
Maxwell-type equations, where electro-magnetic duality is manifest, 
which QCD is a nonabelian gauge theory.

(2)
The dual-superconductor scenario  requires condensation of 
(color-)magnetic monopoles as the key concept, while QCD does not 
have such a monopole as the elementary degrees of freedom.

As the connection between QCD and the dual-superconductor scenario, 
't~Hooft proposed the concept of the {\it abelian gauge fixing,}$^2$
the {\it partial gauge fixing} which only remains 
abelian gauge degrees of freedom in QCD.
By definition, the abelian gauge fixing reduces QCD into 
an abelian gauge theory, where the off-diagonal element of the 
gluon field behaves as a {\it charged matter field} 
similar to $W^\pm_\mu$ in the Standard Model and provides 
a color-electric current in terms of the residual abelian gauge 
symmetry.
As a remarkable fact in the abelian gauge, 
{\it color-magnetic monopoles} appear as {\it topological objects} 
corresponding to the nontrivial homotopy group 
$\Pi_2({\rm SU(N_c)/U(1)}^{\rm N_c-1}) ={\bf Z}^{\rm N_c-1}_\infty$ 
in a similar manner to the GUT monopole.$^{2-4}$
In general, the monopole appears as a topological defect or 
a singularity in a constrained abelian gauge manifold embedded 
in the compact (and at most semi-simple) nonabelian gauge manifold.

Thus, {\it by the abelian gauge fixing, QCD is reduced into an abelian 
gauge theory including both the electric current $j_\mu$ and 
the magnetic current $k_\mu$,} which is expected to provide the 
theoretical basis of the dual-superconductor scheme 
for the confinement mechanism.

\section{Maximally Abelian Gauge and Abelian Projection Rate}  
The abelian gauge fixing is a {\it partial gauge fixing} 
defined so as to diagonalize a suitable gauge-depending variable 
$\Phi [A_\mu (x)]$, and the gauge group $G \equiv {\rm SU}(N_c)_{\rm local}$ 
is reduced into $H \equiv {\rm U(1)}^{N_c-1}$ in the abelian gauge. 
Here, $\Phi [A_\mu (x)]$ can be regarded as a {\it composite Higgs field}
to determine the gauge fixing on $G/H$. 

The {\it maximally abelian (MA) gauge} is a special abelian gauge 
so as to minimize the off-diagonal part of the gluon field, 
$$
R_{\rm off} [A_\mu ( \cdot )] \equiv \int d^4x {\rm tr}
[\hat D_\mu ,\vec H][\hat D^\mu ,\vec H]^\dagger
={e^2 \over 2} \int d^4x \sum_\alpha  |A_\mu ^\alpha (x)|^2
\label{eqn:MAGAUGE}
$$
with the ${\rm SU}(N_c)$ covariant derivative 
$\hat D_\mu \equiv \hat \partial_\mu+ieA_\mu $ and 
the Cartan decomposition 
$A_\mu (x)=\vec A_\mu (x) \cdot \vec H +\sum_\alpha A^\alpha (x)E^\alpha $.
Thus, {\it in the MA gauge, the off-diagonal gluon component 
is forced to be as small as possible by the gauge transformation, 
and therefore the gluon field $A_\mu (x) \equiv A_\mu ^a(x)T^a$ 
mostly approaches the abelian gauge field 
$\vec A_\mu (x) \cdot \vec H$.} 

Since $R_{\rm off}$ is gauge-transformed by $\Omega  \in G$ as 
$$
R_{\rm off}^\Omega 
=\int d^4x {\rm tr}
[\Omega  \hat D_\mu  \Omega ^\dagger, \vec H]
[\Omega  \hat D^\mu  \Omega ^\dagger, \vec H]^\dagger 
=\int d^4x {\rm tr}
[\hat D_\mu ,\Omega ^\dagger \vec H \Omega ]
[\hat D^\mu ,\Omega ^\dagger \vec H \Omega ]^\dagger,
\label{eqn:MAGT}
$$
the MA gauge fixing condition is obtained as 
$$
[\vec H, [\hat D_\mu , [\hat D^\mu , \vec H]]]=0
\label{eqn:MACOND}
$$
>from the infinitesimal gauge transformation of $\Omega $.
In the MA gauge, 
$$
\Phi _{\rm MA}[A_\mu (x)] \equiv [\hat D_\mu , [\hat D^\mu , \vec H]]
\label{eqn:MADIAG}
$$
is diagonalized, and $G \equiv {\rm SU}(N_c)_{\rm local}$ 
is reduced into ${\rm U(1)}^{N_c-1} \times Weyl_{\rm global}$,
where the {\it global Weyl symmetry} is the subgroup of ${\rm SU}(N_c)$ 
relating the permutation of the basis in the fundamental representation. 

In the lattice formalism with the Euclidean metric, 
the MA gauge is defined by maximizing 
the diagonal element of the link variable 
$U_\mu (s) \equiv \exp\{iaeA_\mu (s)\}$, 
$$
R_{\rm diag} [U_\mu ( \cdot )] \equiv \sum_{s,\mu } 
{\rm tr} \{U_\mu (s) \vec H U_\mu (s)^\dagger \vec H \}.
\label{eqn:MALAT}
$$
The ${\rm SU}(N_c)$ link variable is factorized corresponding to the 
Cartan decomposition $H \times G/H$ as 
$$
U_\mu (s)=M_\mu (s)u_\mu (s); \ \ 
M_\mu (s) \equiv \exp\{i\Sigma _\alpha \theta _\mu ^\alpha (s)E^\alpha \}, \ 
u_\mu (s) \equiv \exp\{i \vec \theta _\mu (s) \cdot \vec H \}.
\label{eqn:DECOMP}
$$
Here, the {\it abelian link variable} 
$u_\mu (s) \in H={\rm U(1)}^{N_c-1}$ behaves as 
the abelian gauge field, and the off-diagonal factor 
$M_\mu (s) \in G/H$ behaves as the charged matter field 
in terms of the residual abelian gauge symmetry 
${\rm U(1)}^{N_c-1}_{\rm local}$.
In the lattice formalism, the abelian projection 
is defined by the replacement as 
$$
U_\mu (s) \in G \quad \rightarrow  \quad u_\mu (s) \in H
\label{eqn:APRO}
$$
In the MA gauge, {\it microscopic abelian dominance}
on the link variable is observed as $U_\mu (s) \simeq u_\mu (s)$. 
Quantitatively, in the SU(2) lattice QCD, 
the {\it abelian projection rate} 
$$
R_{\rm Abel} \equiv 
{1 \over 2} \langle {\rm tr} \{ U_\mu (s) u_\mu (s)^\dagger \}
\rangle_{\rm MA} 
={1 \over 2} \langle {\rm tr} M_\mu (s)
\rangle_{\rm MA} \in [0,1]
\label{eqn:APR}
$$
is {\it close to unity in the MA gauge}
for the whole region of $\beta $: it is above 0.88 
even in the strong-coupling limit, where the link variable is 
completely random before the MA gauge fixing.
This is expected to be the basis of 
{\it macroscopic abelian dominance} for the infrared quantities 
as the string tension in the MA gauge.

\section{Abelian Dominance, Monopole Dominance and Global Network of Monopole Current in MA Gauge in Lattice QCD}
Abelian dominance and monopole dominance for NP-QCD 
(confinement, D$\chi $SB, instantons) are 
the remarkable facts observed in the lattice QCD 
in the MA gauge.$^{5-10}$
Abelian dominance means that NP-QCD is described only 
by the ``abelian gluon'', {\it i.e.}, the diagonal gluon component. 
Monopole dominance means that 
the essence of NP-QCD is described only by the 
``monopole part'' of the abelian gluon. 
Here, we summarize the QCD system in the MA gauge 
in terms of abelian dominance, monopole dominance and 
extraction of the relevant mode for NP-QCD.

(a) Without gauge fixing, it is difficult to extract 
relevant degrees of freedom for NP-QCD. 
All the gluon components equally contribute to NP-QCD.

(b) In the MA gauge, QCD is reduced into an abelian gauge theory 
including the electric current $j_\mu $ and the magnetic current $k_\mu $.
The diagonal gluon component (the abelian gluon) 
behaves as the abelian gauge field, 
and the off-diagonal gluon component (the charged gluon) 
behaves as the charged matter field in terms of 
the residual abelian gauge symmetry. 
In the MA gauge, the lattice QCD shows {\it abelian dominance} 
for NP-QCD as the string tension and D$\chi $SB: 
only the abelian gluon is relevant for NP-QCD, 
while off-diagonal gluons do not contribute to NP-QCD. 
In the confinement phase of the lattice QCD, 
there appears the {\it global network of the monopole world-line 
covering the whole system in the MA gauge} as shown in Fig.1.

(c) The abelian gluon can be decomposed into the 
``(regular) photon part'' and the ``(singular) monopole part'', 
which corresponds to the separation of $j_\mu$ and $k_\mu$.$^{6-9}$
The monopole part holds the monopole current $k_\mu $, 
and does not include the electric current, $j_\mu  \simeq 0$. 
On the other hand, the photon part 
holds the electric current $j_\mu $ only, 
and does not include the magnetic current, $k_\mu  \simeq 0$. 
In the MA gauge, the lattice QCD shows {\it monopole dominance} 
for NP-QCD as the string tension, D$\chi $SB and instantons: 
the monopole part leads to NP-QCD, 
while the photon part seems trivial like QED and 
do not contribute to NP-QCD. 

Thus, monopoles in the MA gauge can be regarded as 
the relevant collective mode for NP-QCD, 
and {\it formation of the global network of the monopole world-line 
can be regarded as ``monopole condensation'' in the infrared scale.}
(Of course, local detailed shape of monopole world-lines 
is meaningless for the long-range physics of QCD !)
Hence, the NP-QCD vacuum would be identified as the dual superconductor 
in the infrared scale in the MA gauge.

\section{Origin of Abelian Dominance : Effective Charged-Gluon Mass induced in MA Gauge}

In the MA gauge, only the diagonal gluon component 
is relevant for the infrared quantities 
like the string tension and the chiral condensate, 
and it is regarded as abelian dominance for NP-QCD.
In this section, we study the {\it origin of abelian dominance 
in the MA gauge.}

As a possible physical interpretation, 
abelian dominance can be expressed as {\it generation of the 
effective mass $m_{ch}$ of the off-diagonal (charged) gluon 
by the MA gauge fixing} in the QCD partition functional, 
\begin{eqnarray*}
Z_{\rm QCD}^{\rm MA}&=&\int DA_\mu \exp\{iS_{\rm QCD}[A_\mu ]\} 
\delta (\Phi _{\rm MA}^\pm [A_\mu ])\Delta _{\rm PF}[A_\mu ] \cr\\
&\simeq &
\int DA_\mu ^3 \exp\{iS_{\rm eff}[A_\mu ^3]\}
\int DA_\mu ^\pm \exp\{i\int d^4x \ 
m_{ch}^2 A_\mu ^+A^\mu _- \} {\cal F}[A_\mu ],
\label{eqn:MAQCD}
\end{eqnarray*}
where $\Delta _{\rm FP}$ is the Faddeev-Popov determinant, 
$S_{\rm eff}[A_\mu ^3]$ the abelian effective action 
and ${\cal F}[A_\mu ]$ a smooth functional. 
In fact, if the MA gauge fixing induces the effective mass 
$m_{ch}$ of off-diagonal (charged) gluons, 
the charged gluon propagation is limited within 
the short-range region as $r \lsim m_{ch}^{-1}$, 
and hence off-diagonal gluons cannot contribute 
to the long-distance physics in the MA gauge, which 
provides the origin of abelian dominance for NP-QCD.  

Here, using the SU(2) lattice QCD in the Euclidean metric, 
we study the gluon propagator 
$G_{\mu \nu }^{ab} (x-y) \equiv \langle A_\mu ^a(x)A_\nu ^b(y)\rangle$ 
in the MA gauge 
with respect to the interaction range and strength.$^{11}$ 
As for the residual U(1) gauge symmetry, 
we impose the U(1) Landau gauge fixing 
to extract most continuous gauge configuration and 
to compare with the continuum theory.
In particular, the scalar combination 
$G_{\mu\mu}^a(r)\equiv \sum^4_{\mu=1}\langle 
A_\mu^{~a}(r)A_\mu^{~a}(0)\rangle~(a=1,2,3)$ 
is useful to observe the interaction range of the gluon,
because it depends only on the four-dimensional Euclidean 
radial coordinate $r \equiv (x_\mu x_\mu)^{{1 \over 2}}~$.

We calculate the gluon propagator $G_{\mu \mu }^a(r)$ in the MA gauge 
using the SU(2) lattice QCD 
with $12^3 \times 24$ and $\beta =2.3, 2.35$.$^{11}$
In the MA gauge, 
the off-diagonal (charged) gluon propagates within the 
short-range region $r \lsim 0.4$ fm, 
so that it cannot contribute to the long-range physics, 
although the charged-gluon effect appears at 
the short distance as $r \lsim 0.4$ fm.
On the other hand, the diagonal gluon propagates over the long distance 
and influences the long-range physics.
Thus, we find {\it abelian dominance for the gluon propagator} : 
only the diagonal gluon is relevant at the infrared scale 
in the MA gauge.
This is the {\it origin of abelian dominance for the 
long-distance physics or NP-QCD.}

Since the propagator of the massive gauge boson with mass $M$ behaves as 
the Yukawa-type function $G_{\mu\mu}(r)={3 \over 4\pi^2}
{1 \over r^2}\exp(-Mr)$,
the effective mass $m_{ch}$ of the charged gluon 
can be evaluated from the slope of the logarithmic plot of 
$r^2G_{\mu\mu}^{+-}(r)\sim \exp(-m_{ch}r)$ as shown in Fig.2. 
The charged gluon behaves as a massive particle 
at the long distance, $r \gsim 0.4$ fm.
We obtain the {\it effective mass of the charged gluon} 
as $m_{ch} \simeq 0.9~{\rm GeV}$, 
which provides the {\it critical scale on abelian dominance.} 

\section{Dual Gauge Formalism, Dual Wilson Loop, 
Inter-Monopole Potential and Evidence of 
Dual Higgs Mechanism (Monopole Condensation)}
In this section, we study the dual Higgs mechanism by 
monopole condensation in the NP-QCD vacuum 
in the field-theoretical manner. 
Since QCD is described by the ``electric variable'' as 
quarks and gluons, 
the ``electric sector'' of QCD has been well studied with 
the Wilson loop or the inter-quark potential, however, 
the ``magnetic sector'' of QCD is hidden and still unclear. 
To investigate the magnetic sector directly,
it is useful to introduce the ``dual (magnetic) variable'' 
as the {\it dual gluon} $B_\mu $, 
similarly in the dual Ginzburg-Landau (DGL) theory.$^{3,12}$ 
The dual gluon $B_\mu $ is the dual partner of the abelian gluon 
and directly couples with the magnetic current $k_\mu $.
In particular, 
in the absence of the electric current, 
$\partial_\mu F^{\mu\nu}=j^\nu=0$, 
the dual gluon $B_\mu $ can be introduced 
as the regular field satisfying 
$\partial_\mu B_\nu - \partial_\nu B_\mu={^*\!F}_{\mu\nu}$
and the dual Bianchi identity, 
$
{\partial^{\mu}} {^*\!(}\partial \land B)_{\mu\nu}=0,
$
and therefore 
the argument on monopole condensation becomes transparent.$^{12}$

As was mentioned in Section 3, 
the monopole part in the MA gauge 
and does not include the electric current, $j_\mu \simeq 0$, 
and holds the essence of NP-QCD, which is of interest. 
Then, it is wise to consider the monopole part 
for the transparent argument on the dual Higgs mechanism or 
monopole condensation.
In terms of the dual Higgs mechanism, 
the inter-monopole potential is expected to be short-range 
Yukawa-type, and the dual gluon $B_\mu $ becomes massive 
in the monopole-condensed vacuum.$^{12}$
We define {\it the dual Wilson loop} $W_D$ as the line-integral of 
the dual gluon $B_\mu$ along a loop $C$,
$$
W_D(C) \equiv \exp\{i\oint_C dx_\mu B^\mu \}=
\exp\{i\int\!\!\!\int d\sigma_{\mu\nu}{^*\!F}^{\mu\nu}\},
\label{eqn:DWIL}
$$
which is the {\it dual version of the abelian Wilson loop.}$^{13}$
The potential between the monopole and the anti-monopole 
is derived from the dual Wilson loop as 
$$
V_{M}(R) = -\lim_{T \rightarrow  \infty} {1 \over T}\ln 
\langle W_D(R,T) \rangle.
\label{eqn:DPOT}
$$
Using the SU(2) lattice QCD in the MA gauge,
we study the dual Wilson loop and 
the inter-monopole potential in the monopole part.$^{13}$
The dual Wilson loop $\langle W_D(R,T) \rangle$ seems to 
obey the {\it perimeter law} rather than the area law. 
The inter-monopole potential 
is short ranged and flat in comparison with 
the inter-quark potential as shown in Fig.3.

At the long distance, the inter-monopole potential 
can be fitted by the simple Yukawa potential 
$V_M(r) = -{(e/2)^2 \over 4\pi}{e^{-m_Br} \over r}$.
The dual gluon mass is estimated as 
$m_B \simeq {\rm 0.5GeV}$, 
which is consistent with the DGL theory.$^{3,12}$
{\it The mass generation of the dual gluon $B_\mu $ 
would provide the direct evidence of the dual Higgs mechanism 
by monopole condensation at the infrared scale in the NP-QCD vacuum.}

In the whole region of $r$ including the short distance, 
the inter-monopole potential seems to be fitted by 
the Yukawa-type potential with the effective size 
$R$ of the QCD-monopole as shown in Fig.3. 

The fitting on the global shape of $V_M(r)$ suggests 
the {\it effective monopole size} $R \simeq 0.35{\rm fm}$, 
which would provide the {\it critical scale for NP-QCD in terms of 
the dual Higgs theory as the local field theory.}

\section{Origin of Strong Correlation between Monopoles and Instantons : 
Large Gluon-Field Fluctuation around Monopoles }

There is no point-like monopole in QED, 
because the QED action diverges around the monopole.
The QCD-monopole also accompanies the large fluctuation 
of the abelian action density inevitably. 
In this section, we study the action density 
around the QCD-monopole in the MA gauge 
using the SU(2) lattice QCD.$^{14}$ 

>From the SU(2) plaquette $P^{\rm SU(2)}_{\mu \nu }(s)$ 
and the abelian plaquette $P^{\rm Abel}_{\mu \nu }(s)$, 
we define the ``SU(2) action density'' 
$
S_{\mu \nu }^{\rm SU(2)}(s) 
\equiv 1-{1 \over 2}{\rm tr}P^{\rm SU(2)}_{\mu \nu }(s), 
$
the ``abelian action density'' 
$
S_{\mu \nu }^{\rm Abel}(s) 
\equiv 1-{1 \over 2}{\rm tr}P^{\rm Abel}_{\mu \nu }(s) 
$
and the ``off-diagonal action density'' 
$$
S_{\mu \nu }^{\rm off}(s) 
\equiv S_{\mu \nu }^{\rm SU(2)}(s)-S_{\mu \nu }^{\rm Abel}(s), 
\label{eqn:ACTDOFF}
$$
which is {\it not positive definite.}
In the lattice formalism, 
the monopole current $k_\mu (s)$ is defined on the dual link, 
and there are 12 links around the monopole. 
To investigate the ``local quantity around the monopole'', 
we define the ``local average over the neighboring 12 links 
around the dual link'', 
$$
\bar S(s,\mu) \equiv
{1 \over 12} \sum_{\alpha\beta\gamma} \sum_{m=0}^1 
| \varepsilon_{\mu\alpha\beta\gamma} |
S_{\alpha\beta}(s + m \hat \gamma).
\label{eqn:ACTDL}
$$
>From the total ensemble $\{\bar S(s,\mu )\}_{s,\mu }$, 
we extract the sub-ensemble $\{\bar S(s,\mu )\}_{s,\mu }$ 
{\it around the monopole} in the lattice QCD. 
We show in Fig.4(b) the probability distribution of 
$\bar S_{\rm SU(2)}$, $\bar S_{\rm Abel}$ and 
$\bar S_{\rm off}$ around the monopole in the MA gauge.

{\it Around the monopole in the MA gauge,} 
a {\it large fluctuation} is observed both 
in $S_{\rm Abel}$ and in $S_{\rm off}$, however, 
the total QCD action $\bar S_{\rm SU(2)}$ 
is kept to be small relatively, 
owing to {\it large cancellation} between $\bar S_{\rm Abel}$ and 
$\bar S_{\rm off}$ as shown in Fig.4(b).$^{14}$
Thus, off-diagonal gluons play the essential role to 
appearance of QCD-monopoles 
to keep the total QCD action finite. 
However, in the infrared scale, off-diagonal gluons 
become irrelevant, while 
large abelian-gluon fluctuations originated from 
QCD-monopoles remain to be relevant in the MA gauge. 

Even in the MA gauge, 
off-diagonal gluons largely remain around the QCD-monopole, 
which would provide the {\it effective monopole size} $R$ 
as the critical scale of abelian projected QCD (AP-QCD), 
because off-diagonal gluons become visible and AP-QCD 
should be modified at the shorter scale than $R$, 
which resembles the structure of the 't~Hooft-Polyakov monopole. 
The concentration of off-diagonal gluons around monopoles leads to 
the {\it local correlation between monopoles and instantons}: 
instantons appear around the monopole world-line in the MA gauge, 
because instantons need full SU(2) gluon components 
for existence.$^{7-9,15}$

One of the authors (H.S.) would like to thank 
Professor Y.~Nambu for his useful comments and discussions.

{\bf ~\\ Figure Captions}

Fig.1: The monopole current (the monopole world-line) 
in the MA gauge in the SU(2) lattice QCD with $16^3 \times 4$. 
(a) confinement phase ($\beta =2.2$), 
(b) deconfinement phase ($\beta =2.4$).

Fig.2: 
The scalar correlation $r^2G_{\mu \mu }^a(r)$ 
of the gluon propagator as the function of the 4-dimensional 
distance $r$ in the MA gauge in the SU(2) lattice QCD with 
$12^3 \times 24$ and $\beta =2.3, 2.35.$ 
The charged-gluon propagator behaves as 
the Yukawa-type function, 
$G_{\mu \mu } \sim {\exp(-m_{ch}r) \over r^2}$. 
The effective mass  of the charged gluon 
can be estimated as $m_{ch}\simeq 0.9{\rm GeV}$ 
>from the slope of the dotted line. 

Fig.3: The inter-monopole potential $V_M(r)$ in the monopole part 
in the MA gauge in the SU(2) lattice QCD 
with $16^4$ and $\beta =2.35, 2.4.$ 
$r$ is the 3-dimensional distance 
between the monopole and the anti-monopole. 
The dashed-dotted line denotes the linear part of 
the inter-quark potential in the left figure. 
In the right figure, the dashed curve and the solid curve denote 
the simple Yukawa potential and the Yukawa-type potential with 
the effective monopole size, respectively. 

Fig.4: (a) The total probability distribution $P(\bar S)$ 
and (b) The probability distribution $P_k(\bar S)$ 
around the monopole 
for SU(2) action density $\bar S_{\rm SU(2)}$ (dashed curve), 
abelian action density $\bar S_{\rm Abel}$ (solid curve) 
and off-diagonal part $\bar S_{\rm off}$ (dotted curve) 
in the MA gauge at $\beta =2.4$ on $16^4$ lattice. 
Large cancellation between $\bar S_{\rm Abel}$ and  
$\bar S_{\rm off}$ is observed around the QCD-monopole.

\end{document}